\documentclass[aps,prl,twocolumn,groupedaddress,showpacs,showkeys]{revtex4-2}
\usepackage{graphicx}
\usepackage{dcolumn}
\usepackage{bm}

\begin{document}

\title{Magnetization dynamics and reversal of two-dimensional magnets}
\author{Essa M. Ibrahim}
\author{Shufeng Zhang}
\affiliation{Department of Physics, University of Arizona, 1117 E 4th Street, Tucson, AZ 85721}

\begin{abstract}
Micromagnetics simulation based on the classical Landau-Lifshitz-Gilbert (LLG) equation has long been a powerful 
method for modeling
magnetization dynamics and reversal of three-dimensional (3D) magnets. For two dimensional (2D) magnets, the
magnetization reversal always accompanies the collapse of the magnetization even at the low temperature due to
intrinsic strong spin fluctuation. We propose a micromagnetic theory 
that explicitly takes into account the rapid demagnetization and remagnetization dynamics of 2D
magnets during magnetization reversal. We apply the theory to a single domain magnet to illustrate fundamental differences of
magnetization trajectories and reversal times for 2D and 3D magnets.
\end{abstract}

\date{\today}                     

\maketitle


\section{Introduction}
The classical Landau-Lifshitz-Gilbert (LLG) equation is an essential equation for studying magnetization dynamics driven by an applied magnetic field or spin torques. Almost all of magnetic phenomena related to the magnetic structure and dynamics
can be approximately understood in terms of the LLG equation. The basic assumption in the LLG equation is the invariace of the 
magnitude of magnetization which is a constant at a fixed temperature, independent of the magnetic field or other driving
sources such as the spin torque. As long as 
the temperature is well below the Curie temperature of the 3D magnet, the amplitude of the magnetization is mainly controlled 
by the exchange interaction which is usually several oreders of magntitude larger than the magnetic field or anisotropy field.

For 2D magnets, however, the assumption of a constant amplitude of the magnetization during the dynamic process completely fails at any temperature. To see this, considering a uniaxial anisotropic magnet with the magnetization intially aligned in one of the easy direction denoted as ${\bf m}(0) = m_0 \hat{\bf z}$. When a magnetic field with its magnitude same as the anisotropic field applied in the opposite direction of the magneization, the sum of the magnetic field and the anisotropy field becomes zero.
Consequently, the energy of the quanta of the long wavelength of spin wave excitation, or magnons, scales as $\epsilon_{\bf k} \propto k^2 $ where ${\bf k} $ is a wave vector. Since the magnons are Bosons, it follows that the total number of magnons
$N \propto \int d^2{\bf k} [\exp(\epsilon_{\bf k}/k_BT) -1]^{-1} $ 
diverges at the thermal equalibrium, i.e., the external magnetic field cancels with the anisotropy field and thus the magnon spectrum is gapless, leading to completely demagnization at any finite temperature. 
The above argument is consistent with the Wagner-Mermin's theory which excludes the long-range magnetic ordering for
Heisenberg model without the anisotropy and external fields. Therefore, when the reversal magnetic field reaches the anisotropic field, the magnetization disapears in its initial direction and instead, the magnetization spontaneously appears at the direction of the
reversal magnetic field. We define such collapse of the magnetization in the original direction and re-appeear in the direction of the applied field as demagntization/remagnetization (DMRM).

On the contrary, DMRM does not occur for 3D magnets because the external magnetic field has little effect on the magnitude of the magnetization as long as the magnetic field
is much smaller than the exchange energy and thus the magnetization reversal is governed by the rotation of the local magnetization. In a single domain, the magnetic field must be applied with an angle relative to the magnetization to generate 
the rotation. The rotating dynamics is usually modelled by the classical LLG with the reversal time determined by the
damping parameter and the magnetic field.

The question is how fast is the DMRM process compared with the LLG rotation dynamics? The origin of the DMRM is the change of the magnitude of the
magnetization by the thermal magnons. To see the DMRM time scale, we recall the DMRM experiments in which 3D magnetic films are exposed to a 
short-pulsed laser field, inducing a fast temperature change to the Curie temperature. It has been found the demagnetization is achieved within
a picosecond, and subsequently, the remagnetization follows after the laser field is turned off. The magnetization dynamics
in these experiments involves both longitudinal magnetization dynamics which is the process of reaching the thermal equilibrium of magnons, and the transverse 
magnetization dynamics which is the process of rotation of the order parameter (magntization). The former is several orders of magnitude faster than the latter. 

For magnetization reversal of a 2D magnet, both longitudinal and transverse dynamics are present at any temperature as the 
reversal inevitably invlove the quench of magnetizaion due to magnetization instability generated by the divesgence of the number of magnons. 
In the next Section we propose our model for calculating the magnetization dynamics of 2D magnets.

\section{Dynamic equations for 2D magnets}

To be more specific for the dynamic equations we propose, let's
start with a spin Hamiltonian of the 2D magnet  
\begin{equation}
    {\hat{\cal H}}=-J \sum_{<i,j>}{\hat{\bf S}}_{i} \cdot {\hat{\bf S}}_{j}- A \sum_{<i,j>} \hat{S}_{i}^z \hat{S}_{j}^z-\sum_{i}{\bf H} (t) \cdot \hat{\bf S}_{i}
\end{equation}
where ${\hat{\bf S}}_{i}$ and $\hat{S}_{i}^z$ are respectively the spin and the $z$-component (taken as perpendicular to
the two-dimensional plane) of the spin operators at lattice site ${\bf R}_{i}$, $J$ is the isotropic exchange integral, $A$ is the anisotropic exchange integral, $<ij> $ indicates the sum over nearest neighbors, and $ {\bf H}$ is the time-dependent external field. Before the magnetic field is turned on, the magnetization is initially in the direction of the ferromagnetic ground state. At sufficient low temperature compared to the Curie temperature, we can use the random phase approximation (RPA)
to calculate the average magnetization. The resulting self-consistent equation for the magnetization is
\cite{Tang},
\begin{eqnarray}
M=M_s -\int \frac{d^2k}{(2\pi)^2} \frac{2M}{e^{\beta E_k}-1}
\end{eqnarray}
where $M_s$ is the magnetization at zero temperature, and $E_k$ is the magnon energy; in the long wave length limit,
$E_k =zM( J k^2/2 + 2A) $ where $z$ is the number of nearest-neighbor sites. Equation (2) has a straightforward 
explanation: the magnetization is subtracted by the number of the magnon which is softened by the factor of $M$ at the finite 
temperature. We note that a) Equation (3) is the RPA approximation for spin-1/2, the higher spins would lead a more complicated
equation and the RPA is considered an excellent approximation for temperature far below the Curie 
temperature, and b) we consider the magnetic anisotropy from the anisotropic exchange rather than on-site anisotropy in the form of $-A(S_i^z )^2 $. By using the quadratic dispersion in the energy,  we may integrate out $d^2k$, resulting a simple analytical expression,
\begin{equation}
M=M_s -\frac{1}{ \pi zJ} \left( \frac{1}{\beta} \ln \left| \frac{e^{\beta (\Delta +W)}-1}{e^{\beta \Delta}-1} \right| -W \right)
\end{equation}
where $\Delta = 8AM^2 $ and $W=8\pi JM$ (assuming a square lattice, $z=4$) are the effective magnon gap and the magnon bandwidth, respectively.

At $t=0$, a magnetic field ${\bf H} (t)$ turns on. To determine how the magnetization proceeds with the magnetic field, 
we make the following postulations. First, the longitudinal magnetization is determined by Eq.~(2), in which the band gap is replaced by the total effective field $\Delta = 8AM^2+{\bf M}\cdot {\bf H}(t)$, i.e., 
the magnon distribution reaches the thermal equlibrium much fast than the change of the magnetic field ${\bf H}$. As we stated earlier, this assumption has been supported by the experiments on laser-induced 
magnetization dynamics in which the time scale to reach thermal equilibrium is less than 1 picosecond. When $\Delta$ becomes zero or a negative value at a time $t$, the magnetization will immediately
drops to zero; this is consistent with the Wagner-Mermin theorem. Susequently, the magnetization reappears in the direction of
the total magnetic field. The magnetization switching through this DMRM process is the new physics
for the magnetization reversal. When the gap $\Delta >0$, the solution of Eq.~(3) detrmines the magnitude of the magnetization as a function time. Then our second postulation 
is the transverse dynamic which describes the rotation of the magnetization given by the conventional LLG equation,
\begin{equation}
\frac{d \bf M}{dt}= -\gamma {\bf M} \times  {\bf H}_{eff}+ \alpha \frac{d \bf M}{|{\bf M}|} \times \frac{d \bf M}{dt}
\end{equation}
where $\gamma$ is gyromagnetic ratio, $\alpha $ is the damping parameter, and ${\bf H}_{eff}$ the total effective magnetic field.

The above two hypothesis on the longitudinal and tranverse dynamics completely determine the magnetization dynamics of 2D magnets. The proposed dynmic equations combine the quantum Boson statistics and the classical 
equation of motion. Next, we shall first demonstrate 
how the above dynamics differ from the conventional 3D magnets in a simplest single domain particle.

\section{Application to 2D single domain }

Let us consider a single domain 2D magnet with the initial magnetization aligned in $+{\bf \hat{z}}$. If the magnetic field is applied in the $-{\bf \hat{z}}$ direction, there will be no transverse dynamics without a thermal kick to allow the magnetization deviating
from $\hat{\bf z}$ direction. For a 3D magnet in which the dynamics is always rotational, the time to reverse the magnetization is long even the magnetic field is much larger than the anisotropy field. For 2D, however, as long as the magnetic field is larger than
anisotropy field, the switching to the reversel direction is considered immediately via the demagnetization and remagnetization in the direction of the magnetic field. More interesting case is the reversal magnetic field in a direction not parallel to the $-{\bf \hat{z}}$
direction. As an example, we consider the magnetic field at $45^o$ with respect to $-{\bf \hat{z}}$. In this case, the magnetization trajectories display several distinct behaviors as shown in Fig.~(1). For a small magnetic field, 
magnetization dynamics is governed by the rotation without reversal and the tranjectories are almost identical for 2D and 3D magnets (except that the amplitude of the magnetization has a small variation in the trajectories).

When the external field reaches a critical value $H_c$, the magnetization trajectories for 2D and 3D magnets qualitatively differ. In 3D, the magnetization processes with spiral rotation towards the final equlibrium position as shown in Fig.~2(h) while for 2D, the
magnetization starts rotation for a short period of time before the gap $\Delta $ becomes zero, at which time the demagnetization occurs and the magnetization appears at the direction of the field. After the remagnetization, the anisotropy field will make the magnetization rotates to the final equilibrium direction determined by the competition between the external field at $45^o$ and the anisotropy field at ${-\hat{\bf z}}$, as shown in Fig.~2(c). Thus, the magnetization reversal has three distinct processes: rotation, DMRM, and the final rotation. 

If one further increases the external magnetic field, the magnon energy becomes negative for the initial state of the magnetization. In this case, the swithing immediately occurs via DMRM without the prior rotation. The final rotation characterizes the 
transverse dynamics that descrbs the rotation of the magnetization from parallel to ${\bf H}$ to the total field direction, as noted earlier.

An important measure for magnetic memory elements is the switching time which is defined as the time $t_s$ for the magnetization crossing the ``equater" from the initial position. For the 3D magnets, the time scale is limited by the product of the
damping parameter and the applied field, typically of the order of sub-nanoseconds to a few nanoseconds. The switching time could be much shorter with the DMRM process of the 2D magnets. In Figure x, we show the swicthing
time as the function of the applied field. If the field is too small, both 2D and 3D are unable to switch the magnetization. At an intermediate field, the 2D magnet switching is much faster because the magnetization rotation is only a small portion of the reversal projectory before the DMRM process. At a sufficiently large field, the 2D magnet switching is instantaneous, only limited by the relaxation mechanism of thermal magnons. 

As the DMRM process depends on the temperature, we show temperature dependence of the switching time in Fig.xx.---discussion on the figure follows.

\section{Discussion and Conclusion}

We have proposed the magnetization dynamic equations by considering the longitudinal and transverse relaxations. The logitudinal process is governed by the quantum statistics, i.e., Boson statistics of magnons. We postulate that the equalibrium distribution
of the magnons for a given spin Hamiltonian is much faster than the classical transverse dynamics. Our proposed quantum-classical magnetization dynamic processes introduce a novel demagnetization and remagnetization dynamics for 2D magnets which is the key
physics that makes the magnetization reversal much faster. Our example for a single domain gives quantitative different behaviors for 2D and 3D magnets.

The calculation scheme we proposed here shall replace the convensional micromagnetics which do not consider the DMRM physics. Since the DMRM is fundamentally present in 2D magnets as shown in Wagner-Menmin theory, any micromagnetic
calculations on 2D materials must take into account DMRM dynamics.

\begin{figure}
    \centering
    \includegraphics[width=8.6 cm]{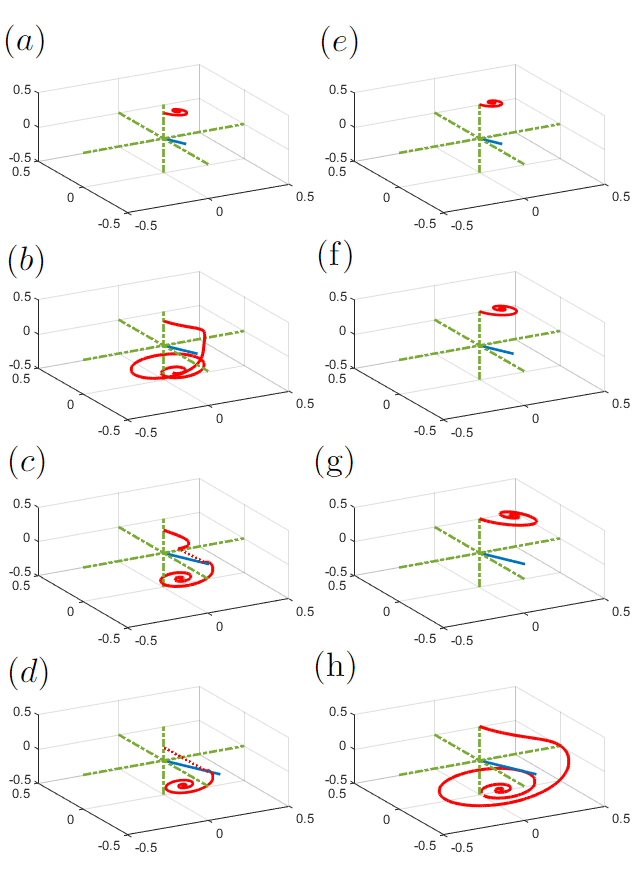}
    \caption{Comparison between the magnetization trajectories for the 3D case (Right) and the 2D case (Left) at different intensities of the external magnetic field (from the top to the bottom $H_{ex}= 0.2J, 0.3J, 0.4J , 0.5J$). The dotted line represents the process of DMRM and the blue line represents the external magnetic field.}
    \label{Figure 1}
\end{figure}

\begin{figure}
    \centering
    \includegraphics[width=8.6 cm]{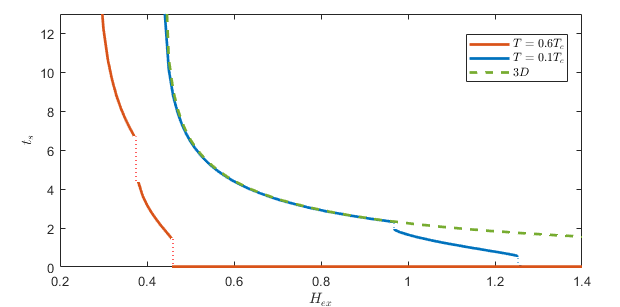}
    \caption{The switching time for different intensities of external magnetic field. Starting from a threshold value of $H_{ex}$, we find that increasing the external field continuously reducing the switching time until the point when the field is strong enough to make the energy gap collapses before the smooth switching process finishes, which in turn reduces the switching time even more. And after $H_{ex}>2zAM_s(T)$, the DMRM process happens instantaneously. We can also notice that as the temperature decreases, the behaviour reduces to the classical 3D known behaviour. For this graph we used $z=4$, $A=0.25J$,and $\alpha=0.2$.  }\label{Figure 2}
    
\end{figure}

This work was partially supported by the U.S. National Science Foundation under Grant No. ECCS-2011331.


\end{document}